\documentclass[11pt,twocolumn,twoside,a4paper,amsmath,amssymb,aps,showkeys,showpacs]{revtex4}

\usepackage{amsfonts}
\usepackage{graphics} 
\usepackage{epsfig}
\usepackage{fancyheadings}

\begin{document}
\thispagestyle{myheadings}
\rhead[]{}
\lhead[]{}
\chead[Radchenko N.V.]{Difference in multiplicity distributions in $pp$ and $p\bar{p}$ at high energies}

\title{Difference in multiplicity distributions in proton-proton and
proton-antiproton collisions at high energies}

\author{Radchenko N.V.}

\email{nvrad@mail.ru}

\affiliation{Novgorod State University, Novgorog-the-Great,
Russia}

\begin{abstract}
Secondary charged hadrons multiplicity distributions in
proton-proton and proton-antiproton collisions differ on
principle. There are three types of inelastic processes in
proton-antiproton scattering. The first type is production of
secondary hadrons shower at gluon string decay. The second type is
shower produced from two quark strings decay, the third type is
shower produced from three quark strings decay. At the same time
there are only two types of inelastic processes for proton-proton
scattering - gluon string shower and two quark strings shower.
Theoretical description of multiplicity distributions is obtained
for proton-proton collisions at energies from 44.5 GeV to 200 GeV
and for proton-antiproton collisions at energies from 200 GeV to
1800 GeV. The difference between proton-proton and
proton-antiproton multiplicity distributions is discussed. The
predictions of multiplicity distribution and mean multiplicity at
LHC energy are given.\end{abstract}



\thanks{Author is very grateful to Organizing Committee for financial support and hospitality}

\maketitle


\section{Introduction}
Multiplicity distributions and total cross sections are the main
characteristics of hadrons multiple production at high energies.
Analysis of secondary hadrons multiplicity distributions is very
important for verification of different phenomenological
approaches and models. Experimental data on multiplicity
distribution are highly informative and can be measured with
sufficient accuracy which gives significant advantage when
comparing with theoretical calculations.

\section{Low Constituents Number Model and total cross sections}
We are based on the Low Constituents Number Model (LCNM) which was
proposed by V.A.~Abramovsky and O.V.~Kancheli in 1980~\cite{bib1}.
Three steps are concretized in this model: preparation of
colliding hadrons initial state; interaction; moving apart of
reaction products.

1) On the first step before the collision there is small number of
constituents in hadrons. They are  valence quarks and few gluons
which fill in the whole spectrum in rapidity space.

2) On the second step the hadrons interaction is carried out by
gluon exchange between the valence quarks and initial gluons and
the hadrons gain the color charge.

3) On the third step after interaction the colored hadrons move
apart and when the distance between them becomes larger than the
confinement radius, the lines of color electric field gather into
the string. This string breaks out into secondary hadrons.

In the model total cross sections of proton-proton and
proton-antiproton collisions are described by the following
formulae
\begin{eqnarray}
\label{1} \sigma_{tot}^{p(\bar{p})p}=63.52s^{-0.358}\mp
35.43s^{-0.56} +\nonumber\\+\sigma_0+ \sigma_1\ln s+\sigma_2(\ln
s)^2+\nonumber\\+\sigma_3(\ln s)^3+[\mbox{higher powers of $\ln
s$}].
\end{eqnarray}
where first two terms present non vacuum contributions and were
taken from the paper by Cudell et. al.~\cite{bib2}. There are also
constant term and terms proportional to powers of logarithm of
full energy. Constant term corresponds to gluon exchange between
initial states containing only valence quarks. Powers of logarithm
are equal to number of gluons in initial states of hadrons.

The fitting of formulae~(\ref{1}) to total cross sections of $pp$
and $p\bar{p}$ collisions (Fig.1, data were taken
from~\cite{bib3}) showed that there are only two gluons in initial
state. Contribution of the third gluon is about 1~mb at energy
14~TeV and thus is negligible. The values of parameters are
$\sigma_0=20.08\pm0.42$, $\sigma_1=1.14\pm0.13$,
$\sigma_2=0.16\pm0.01$.

\begin{figure}
\includegraphics[scale=0.43]{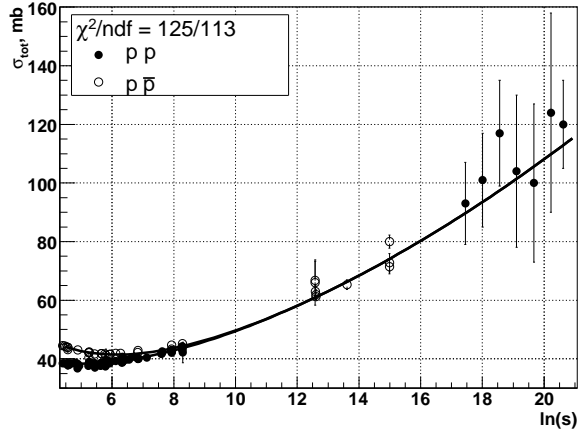}
\caption{Fitting of total cross sections of $pp$ and $p\bar{p}$
collisions in LCNM}
\end{figure}

Having limited number of gluons we can describe the types of
inelastic processes in proton-proton and proton-antiproton
interactions.

\section{Three types of inelastic processes}
From the obtained result of presence of only two gluons in initial
state it follows existence of three types of inelastic processes
in $pp$ and $p\bar{p}$ collisions.

There are three types of inelastic processes in proton-antiproton
interaction. The first type is production of hadrons shower from
decay of gluon string, it corresponds to constant contribution to
total cross sections (Fig.~2a). The second type is shower produced
from decay of two quark strings, it corresponds to contributions
from one and two gluons (Fig.~2b). The third type is shower
produced from decay of three quark strings, it corresponds to part
of contribution from  two gluons (Fig.~2c). In this case quark
strings are produced between every quark of proton and antiquark
of antiproton.

\begin{figure}
\includegraphics[scale=0.6]{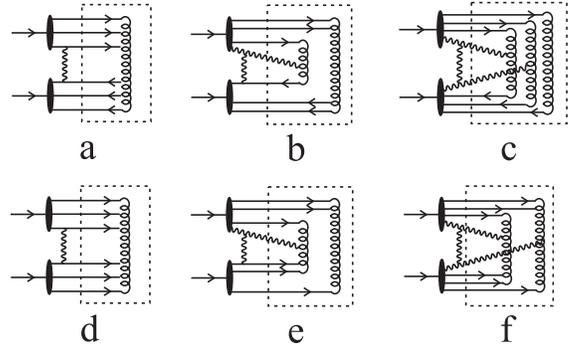}
\caption{Types of inelastic processes for $p\bar{p}$ (a -- c) and
$pp$ interaction (d -- f). Solid lines correspond to quarks and
antiquarks, wavy lines to gluons, spirals to strings. Interaction
in final state is marked by dashed block.}
\end{figure}
\begin{figure}
\includegraphics[scale=0.6]{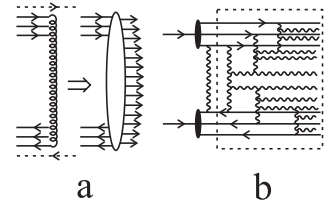}
\caption{Hadrons production in gluon string. a) Amplitude of
transition of two octet states produced after gluon exchange to
hadrons, dashed lines show direction of octet states. b) One of
possible diagrams of Fig.~2a; dashed block shows one of possible
diagram of amplitude of transition of two octet states of Fig.~2a
to hadrons}
\end{figure}

At the same time there are only two types of inelastic processes
in proton-proton interaction, they are shower from gluon string
(Fig.~2d) and shower from two quark strings (Fig.~2e,f). There are
no three quark strings in $pp$ interaction because in this case
strings can be formed between quark of one proton and diquark of
another proton.

\section{Multiplicity distributions for different processes}
We suppose that hadrons multiplicity distribution in gluon string
is normal distribution (Fig.~3).

Gluon pairs are sequentially produced in final state marked by
dashed block. This production takes place until energy of pairs
approaches hadrons mass. Then gluon pairs decays at several
observable hadrons.

Every diagram of this type gives certain number of secondary
hadrons. Number of such diagrams is considerably large. There is
no suppression on energy because exchange is of vector type.
Contributions from this diagrams have the same order of magnitude.
So secondary hadrons multiplicity as random variable has to obey
normal distribution because of central limit theorem of
probability theory. Thus multiplicity distribution in gluon string
is normal distribution.

We suppose that hadrons multiplicity distribution in quark string
is negative binomial distribution (NBD). This is in good agreement
with data on $e^+e^-$ annihilation at low energies where one quark
string is produced with assurance. So multiplicity distribution in
two quark strings is convolution of two NBD, namely NBD with
double parameters, and in three quark strings it is convolution of
three NBD, namely NBD with triple parameters.

Experimental charged multiplicities are normalized on non single
diffraction cross sections,
$\sigma_{nsd}=\sigma_{tot}-\sigma_{el}-\sigma_{sd}$. Pomeron
contributions are the same as for total cross sections
\begin{equation}\label{2}
\sigma^{nsd}_{vac}=\sigma_0^{nsd}(1+\delta_1^{nsd}\ln
s+\delta_2^{nsd}\ln^2 s),
\end{equation}
here we divided formulae by constant term for convenience.

The fitting of vacuum contributions to non single diffraction
cross sections is shown on Fig.~4, data were taken
from~\cite{bib4, bib5}. The value of $\chi^2$ to number of degrees
of freedom 14 to 8 is caused by difference between data of UA4 and
UA5 collaborations.

\begin{figure}
\includegraphics[scale=0.43]{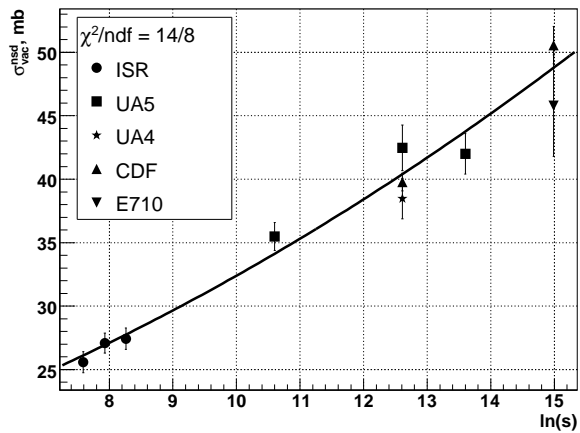}
\caption{Vacuum contributions to non single diffraction cross
sections of $pp$ and $p\bar{p}$ collisions}
\end{figure}

The obtained coefficients $\delta_1^{nsd}=0.075\pm0.011$ and
$\delta_2^{nsd}=0.007\pm0.001$ determine the weights of
configurations with gluon string, two quark strings, three quark
strings.

Weight of normal distribution is defined by formulae~(\ref{3}) and
it is the same for $pp$ and $p\bar{p}$ collisions.
\begin{equation}\label{3}
\frac{1}{1+\delta_1^{nsd}\ln s+\delta_2^{nsd}(\ln s)^2}
\end{equation}

Weights of double NBD are different for $pp$~(\ref{4}) and
$p\bar{p}$~(\ref{5}) interactions because in case of
proton-antiproton configuration with two gluons also gives
contribution to three quark strings. It is taken into account by
coefficient $c$, which does not depend on energy of collision.
\begin{equation}\label{4}
\frac{\delta_1^{nsd}\ln s+\delta_2^{nsd}(\ln
s)^2}{1+\delta_1^{nsd}\ln s+\delta_2^{nsd}(\ln s)^2}
\end{equation}

\begin{equation}\label{5}
\frac{\delta_1^{nsd}\ln s+(1-c)\,\delta_2^{nsd}(\ln
s)^2}{1+\delta_1^{nsd}\ln s+\delta_2^{nsd}(\ln s)^2}
\end{equation}

Therefore weight of triple NBD in configuration with three quark
strings in $p\bar{p}$ collision is defined by formulae~(\ref{6}).
\begin{equation}\label{6}
\frac{c\,\delta_2^{nsd}(\ln s)^2}{1+\delta_1^{nsd}\ln
s+\delta_2^{nsd}(\ln s)^2}
\end{equation}
But coefficient $c$ can not be evaluated from the existing data,
so we consider two possible values of it --  0.25  and 0.75.

\section{Experimental data on charged multiplicity distributions}
With assumptions stated in previous sections we described
experimental data on charged particle multiplicity for $pp$
collisions at energy range from  $\sqrt{s}=44.5$ to
200~GeV~\cite{bib5, bib6}  and for $p\bar{p}$ collisions from 200
to 1800 GeV~\cite{bib7} -- \cite{bib9}.

In case of $pp$ interaction weights of distributions are strictly
defined by formulaes~(\ref{3}), (\ref{4}). In Fig.~5 multiplicity
distributions in $pp$ interaction at $\sqrt{s}=44.5$  and
$\sqrt{s}=62.2$~GeV are shown. Here dash-dot line presents normal
distribution in gluon string and dotted line presents double NBD
in two quark strings, solid line is sum of these distributions.

\begin{figure}
\includegraphics[scale=0.43]{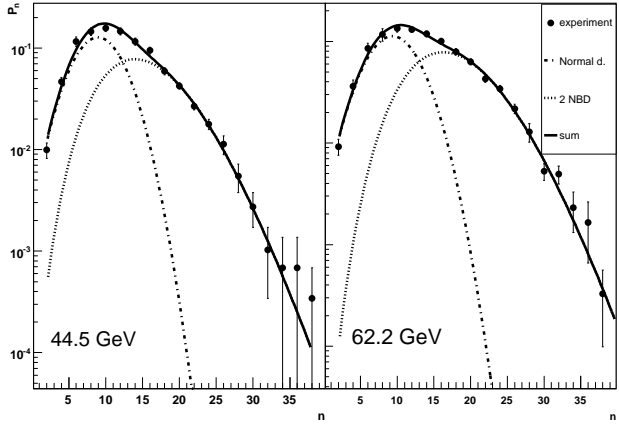}
\caption{$pp$ interaction at $\sqrt{s}=44.5$ ($\chi^2/ndf=13/15$)
and $\sqrt{s}=62.2$~GeV ($\chi^2/ndf=13/16$)}
\end{figure}
\begin{figure}
\includegraphics[scale=0.43]{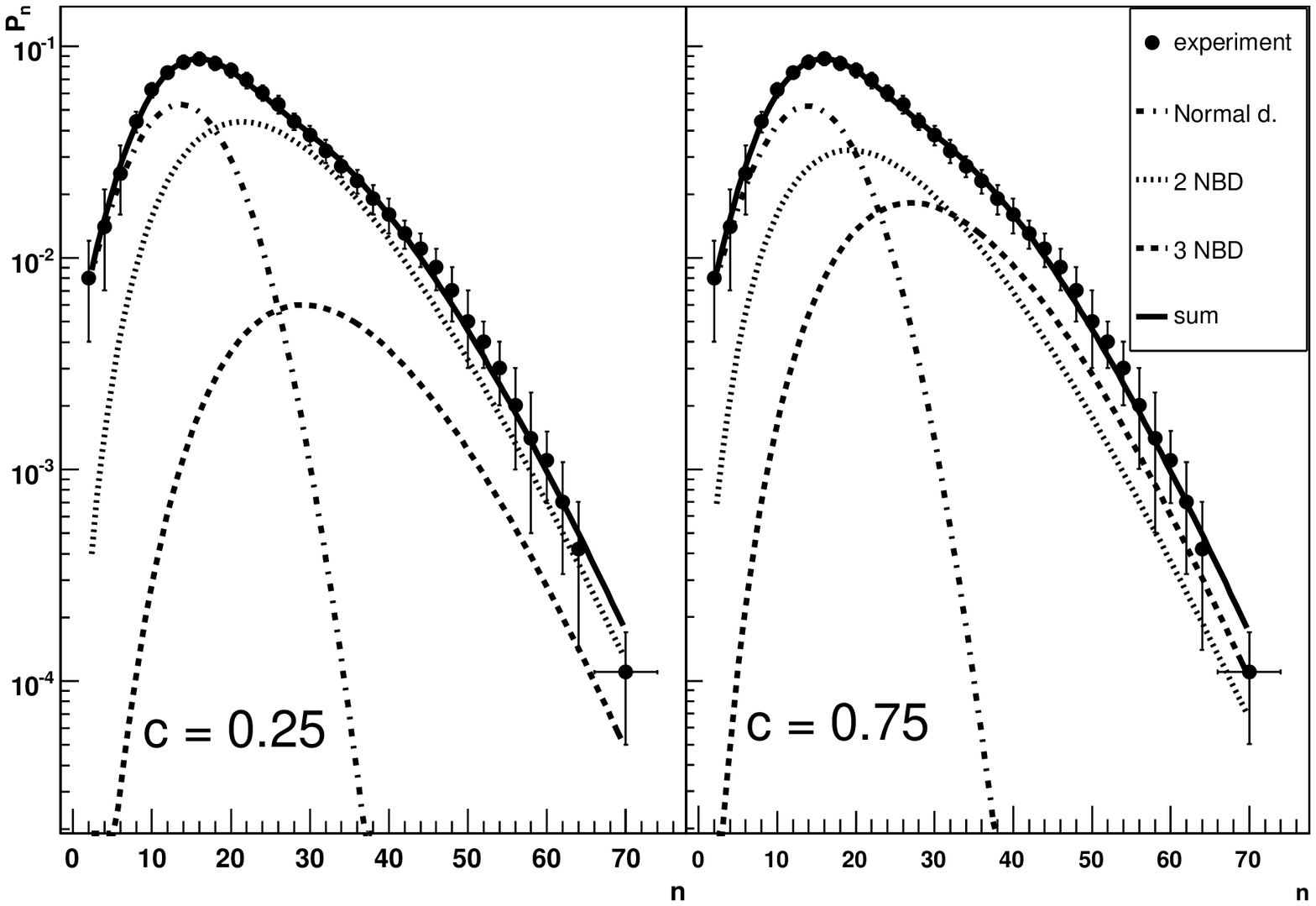}
\caption{$p\bar{p}$ interaction at $\sqrt{s}=200$~GeV}
\end{figure}
\begin{figure}
\includegraphics[scale=0.43]{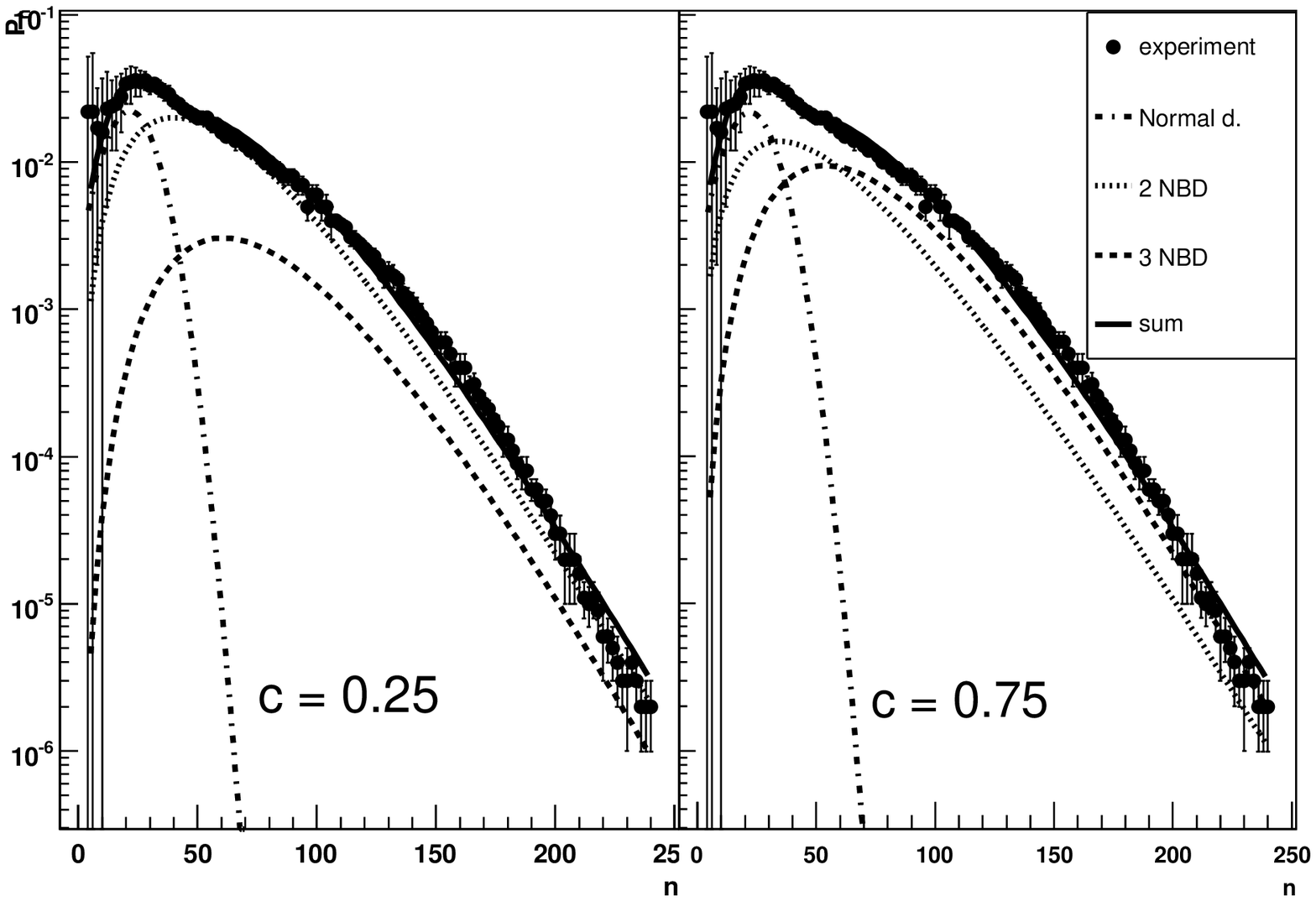}
\caption{$p\bar{p}$ interaction at $\sqrt{s}=1800$~GeV}
\end{figure}
\begin{figure}
\includegraphics[scale=0.43]{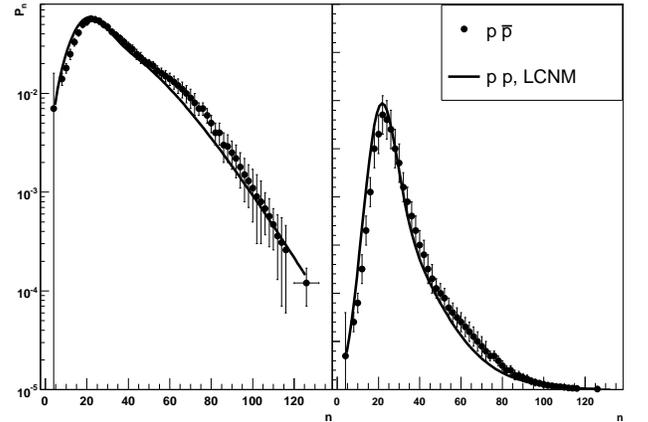}
\caption{$p\bar{p}$ interaction at $\sqrt{s}=900$~GeV (points) and
prediction for $pp$ at the same energy in LCNM for $c=0.75$ (line)
in logarithmic and linear scale}
\end{figure}

But in case of $p\bar{p}$ interaction we have additional
configuration with three quark strings, so weights of
distributions are defined by formulaes~(\ref{3}), (\ref{5})
and~(\ref{6}). The last two formulaes have undefined parameter --
coefficient $c$ which describes appearance of three quark strings.

In Fig.~6 two fittings of $p\bar{p}$ collision at
$\sqrt{s}=200$~GeV are shown. Here dash-dot line presents normal
distribution in gluon string, dotted line presents double NBD in
two quark strings and dashed line presents triple NBD in three
quark strings, solid line is sum of these distributions. On the
left side the case when 25 per cent of two gluons give three quark
strings is shown, on the right side -- the case when 75 per cent
of two gluons give three quark strings. Values of $\chi^2$ to
number of degrees of freedom are the same for both cases,
$\chi^2/ndf=4/29$. As energy increases higher value of coefficient
$c$ gives slightly better value of $\chi^2/ndf$, for example, at
the highest available energy  $\sqrt{s}=1800$~GeV
$\chi^2/ndf=138/115$ when $c=0.25$ and $\chi^2/ndf=126/115$ when
$c=0.75$ (Fig.~7).

After fitting all data we obtained energy dependence of parameters
of gluon and quark strings ad thus we can predict the multiplicity
distribution for $pp$ collisions at 900~GeV. This is the first
high enough energy for which it will be possible to compare
multiplicity distributions for $pp$ and $p\bar{p}$ interactions
directly after start up of LHC. In the case of $c=0.25$  the
shapes for both reactions are practically the same. But in the
case of $c=0.75$ the systematic difference may be observed in
region of ``shoulder'' given very precise measurements (Fig.~8).

\section{Predictions for LHC energy 14 TeV}
We predict the value of total cross section at $\sqrt{s}=14$~TeV
to be
$$
\sigma_{tot}^{pp}=101.30\pm6.65.
$$

\begin{figure}
\includegraphics[scale=0.43]{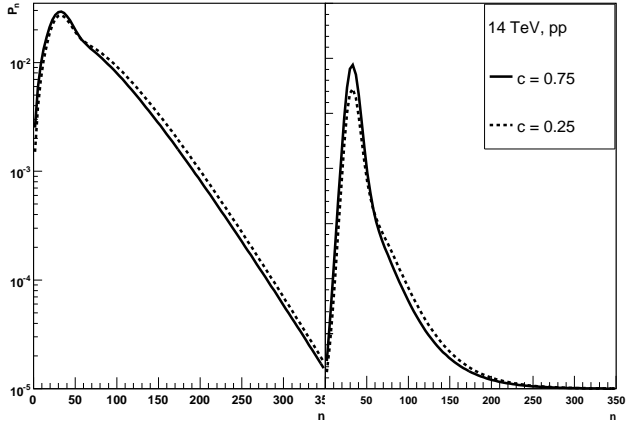}
\caption{Prediction of multiplicity distribution for $pp$
interaction at $\sqrt{s}=14$~TeV in LCNM for $c=0.75$ (solid line)
and for $c=0.25$ (dashed line) in logarithmic and linear scale}
\end{figure}

The multiplicity distribution for $pp$ interaction can be
described as sum of normal distribution in gluon string and double
negative binomial distribution in two quark strings. Weights of
these distributions are strictly defined by formulaes~(\ref{3}),
(\ref{4}). Parameters of normal distribution are mean multiplicity
in gluon string $\bar{n}_{gl}$ and standard deviation
$\sigma_{gl}$. Parameters of double NBD are mean multiplicity in
quark string $\bar{n}_q$ and shape parameter $k_q$. Since
parameters of quark and gluon strings must be obtained at
simultaneous fitting of $pp$ and $p\bar{p}$ distributions, they
contain uncertainty of three quark strings coefficient $c$. So we
have two sets of energy dependencies for gluon and quark string
parameters.

$$\begin{array}{l}
\underline{c=0.25}\\[1mm]
\bar{n}_q=3.45+0.46\ln^2\sqrt{s}\\[1mm]
k_q^{-1}=-0.24+0.11\ln\sqrt{s}\\[1mm]
\bar{n}_{gl}=-5.86+3.77\ln\sqrt{s}\\[1mm]
\sigma_{gl}=-2.67+1.52\ln\sqrt{s}\\[1mm]
\underline{c=0.75}\\[1mm]
\bar{n}_q=3.73+0.41\ln^2\sqrt{s}\\[1mm]
k_q^{-1}=-0.30+0.13\ln\sqrt{s}\\[1mm]
\bar{n}_{gl}=-6.65+3.97\ln\sqrt{s}\\[1mm]
\sigma_{gl}=-2.75+1.55\ln\sqrt{s}
\end{array}
$$

Predictions for charged multiplicity distribution for energy
$\sqrt{s}=14$~TeV are shown in Fig.~9 for both cases of three
quark strings coefficient $c$. The values of mean multiplicity are
the following
$$
\bar{n}_{pp}=  \left\{\begin{array}{l}68.59\pm4.47,\quad\mbox{if
$c=0.25$;}\\ 63.20\pm4.19,\quad\mbox{if
$c=0.75$.}\end{array}\right.
$$

It will be possible to distinguish between two values of
coefficient $c$ when experimental data are available.



\end{document}